\documentclass{article}

\usepackage{arxiv}
\usepackage{amsmath}
\usepackage[utf8]{inputenc} % allow utf-8 input
\usepackage[T1]{fontenc}    % use 8-bit T1 fonts
\usepackage{hyperref}       % hyperlinks
\usepackage{url}            % simple URL typesetting
\usepackage{booktabs}       % professional-quality tables
\usepackage{amsfonts}       % blackboard math symbols
\usepackage{nicefrac}       % compact symbols for 1/2, etc.
\usepackage{microtype}      % microtypography
\usepackage{lipsum}
\usepackage{graphicx}
\usepackage{xcolor}
\usepackage{verbatim}
\graphicspath{ {./images/} }

\title{Non-linear shrinkage of the price return covariance matrix is far from optimal for portfolio optimisation }

\author{
 Christian Bongiorno \&
Damien Challet \\
Université Paris-Saclay, CentraleSupélec\\
Laboratoire de Mathématiques et Informatique pour la Complexité et les Systèmes\\
  91192 Gif-sur-Yvette, France
}

\begin{document}
\maketitle

\begin{abstract}
Portfolio optimization requires sophisticated covariance estimators that are able to filter out estimation noise. Non-linear shrinkage is a popular estimator based on how the Oracle eigenvalues can be computed using only data from the calibration window. Contrary to common belief, NLS is not optimal for portfolio optimization because it does not minimize the right cost function when the asset dependence structure is non-stationary. We instead derive the optimal target. Using historical data, we quantify by how much NLS can be improved. Our findings reopen the question of how to build the covariance matrix estimator for portfolio optimization in realistic conditions.
\end{abstract}





\section{Introduction}

Covariance filtering is essential in multivariate Finance \cite{michaud1989markowitz} and in all scientific fields (see \cite{bun2017cleaning} for a recent review).
A very popular  family of estimators filters the covariance matrix by only changing its eigenvalues while keeping its eigenvectors untouched. They are known as  Rotationally Invariant Estimators (RIEs). The best known RIE  is linear shrinkage \cite{ledoit2004well}. More recently, several methods of non-linear shrinkage (NLS) have been introduced \cite{ledoit2011eigenvectors,ledoit2012nonlinear,bartz2016cross,bun2016rotational}. 
 NLS asymptotically converges to the Oracle estimator, which minimizes the Frobenius  (element-wise square) distance between the filtered and the true population covariance matrix. The asymptotic limit corresponds to infinitely large data matrices at fixed aspect ratio (the number of lines divided by the number of columns). Two important conditions must be met to achieve this convergence: the dependence structure between the time series must be constant, and the price return distribution  must not be too heavy-tailed   \cite{ledoit2011eigenvectors,ledoit2012nonlinear,bartz2016cross,bun2016rotational}.

Because NLS is provably a best estimator is some sense, it is also widely used for portfolio optimization (\cite{BODNAR2018371,DING2021502,rubio2012performance,yang2015robust,yang2014minimum,jrfm12010048,zhao2021risk} among hundreds of references), including in the much cited the state-of-the-art combination of dynamic conditional covariance and NLS \cite{engle2019largecov, de2022large}. This results from the implicit belief that Frobenius-optimal eigenvalues are also optimal for portfolio optimization, or equivalently that the Oracle eigenvalues inevitably provide the best (or almost the best) eigenvalues for portfolio optimization. There is however no proof that it is actually the case. 

Our main hypothesis is that the covariance matrix eigenvalues that are best for global minimum variance minimization do not coincide with the Oracle eigenvalues and thus that non-linear shrinkage is not optimal in general. A conceptually simple way to demonstrate this hypothesis is to compute the RIE eigenvalues that do yield the optimal Global Minimum Variance portfolio and show that this RIE outperforms the Oracle. Such an estimator is obtained solving a quadratic programming problem, which admits an optimal solution different from the Oracle.

We investigate with real financial data where this discrepancy comes from and reversely in which conditions the NLS is a good approximation to the optimal eigenvalues. We found that only when both calibration and test windows are very large with respect to the number of stocks and when the covariance coefficients between calibration and test differs only due to sample size noise (in other words, both sample covariance matrices have the same expected covariance matrix), then the two estimators share similar performances.

%To characterize the conditions where the discrepancy between the Oracle and the optimal estimator are marginal, and NLS is applicable, we numerically explored real financal time-series.

%To check our second hypothesis,  we  stationarize the data and find that our first result is due to non-stationary covariance matrices. We finally check that  NLS is optimal in the small-dimension limit, which corresponds to the unrealistic limit of large in-sample and out-of-sample windows with a small, fixed, number of assets.
 
% We only study GMV portfolios instead of more generic portfolio optimization problems that involve return expectations in order to focus on the influence of the covariance matrix filtering, leaving the problem of return expectations/prediction aside. 

%A further hypothesis that this is because the Oracle eigenvalues themselves are suboptimal as the Frobenius distance is not the appropriate cost function when the dependence structure between financial assets depends on time. 
\section{Problem Statement}
Consider $n$ price return time series split into a calibration time window of length  $\delta_{in}$ and a test time windows of length $\delta_{out}$,  and let us denote the in-sample (empirical) covariance matrix by ${\bf \Sigma}^\textrm{in} \in \mathbb{R}^{n \times n}$ and the out-of-sample (realized) covariance matrix by ${\bf \Sigma}^\textrm{out} \in \mathbb{R}^{n \times n}$ respectively. According to the spectral theorem, the in-sample covariance matrix can be decomposed into a sum of terms involving its eigenvalues $\mathbf{\lambda}=(\lambda_k)\in\mathbb{R}^n$ and their associated eigenvectors components $v_{i}=(v_{ik})\in\mathbb{R}^n$ as
\begin{equation}
    \Sigma_{ij}^\textrm{in} = \sum_{k=1}^n \lambda_k^\textrm{in} v_{ik}^\textrm{in}v_{jk}^\textrm{in}.
\end{equation}
An RIE estimator $\Xi({\bf \lambda}^*)$ uses filtered eigenvalues, which yields the spectral decomposition
\begin{equation}
    \Xi({\bf \lambda}^*)_{ij} = \sum_{k=1}^n \lambda_k^* v_{ik}^\textrm{in}v_{jk}^\textrm{in},
\end{equation}
where $\bf{\lambda}^*$ are a set of filtered eigenvalues obtained with some procedure and ${\bf v}^\textrm{in}$ are still the eigenvectors of ${\bf \Sigma}^\textrm{in}$.

We aim to compare the filtered eigenvalues from  NLS and from the ones that are actually optimal for portfolio optimization. 

\subsection{Oracle Eigenvalues}
The Oracle eigenvalues are obtained from the out-of-sample covariance matrix from 
\begin{equation}
    \lambda^\textrm{Oracle}_k = \sum_{i=1}^n \sum_{j=1}^n v_{ik}^\textrm{in} \Sigma_{ij}^\textrm{out} v_{jk}^\textrm{in},
\end{equation}
where $v^\textrm{in}$ are the in-sample eigenvectors. 

Such an eigenvalue correction provably produces an estimator that minimizes the Frobenius norm \cite{ledoit2011eigenvectors,bun2016rotational}
\begin{equation}
    ||\Xi(\lambda^\textrm{Oracle}) - \Sigma^\textrm{out}||_F = \sum_{i=1}^n\sum_{j=1}^n \left[ \Xi(\lambda^\textrm{Oracle})_{ij} - \Sigma^\textrm{out}_{ij} \right]^2.
\end{equation}

\subsection{Optimal RIEs for GMV Portfolios}

The simplest portfolio optimization problem (and the most relevant one to assess covariance filtering methods)  is  Global Minimum Variance (GMV) portfolios. We denote the fraction of capital assigned to each possible asset $i=1,\cdots,n$ by $\mathbf{w}\in \mathbb{R}^n$. GMV portfolios aim to minimize the realized portfolio variance $\Sigma^\textrm{out}$ at fixed net leverage $\sum_i w_i=1$. Mathematically, the problem can be written as the function of the weights as
\begin{equation}\label{eq:minSout}
    \min_{w} \sum_{i=1}^n \sum_{j=1}^n w_i \Sigma_{ij}^\textrm{out} w_j~~~\textrm{with }\sum_{k=1}^N w_k=1.
\end{equation}
This problem is readily solved if the future is known: the optimal GMV weights are given by
\begin{equation}
    \mathbf{w}_\textrm{opt}=\frac{(\Sigma^\textrm{out})^{-1}\mathbf{e}}{\mathbf{e}'(\Sigma^\textrm{out})^{-1}\mathbf{e}},
\end{equation}
where  $\mathbf{e}={1,\cdots,1}$ is a $n$-dimensional vector of ones.

The main contribution of this paper is to show how not optimal  NLS is for GMV portfolio optimization in a practical context. To this end, we compute the GMV-optimal RIE, which constrains the weights $\mathbf{w}$ to be written as a function of $\Xi(\mathbf{\lambda}^*)$ instead of $\Sigma^\textrm{out}$, the optimization variables being  $\Xi$'s eigenvalues. The optimal weights are now
\begin{equation}\label{eq:GMV}
    w_k = \frac{\sum_{j=1}^n \Xi_{kj}^{-1}}{\sum_{i=1}^n\sum_{j=1}^n \Xi_{ij}^{-1}},
\end{equation}
where ${\bf \Xi}^{-1}$ is the inverse covariance matrix RIE  whose spectral decomposition is
\begin{equation}\label{eq:invert}
    \Xi_{ij}^{-1} = \sum_{k=1}^n \frac{1}{\lambda_k^*} v_{ik}^\textrm{in}v_{jk}^\textrm{in}.
\end{equation}

It is important to point out that the denominator of \ref{eq:GMV} is a normalization factor which ensures that the sum of the weights equals one. Equation \ref{eq:GMV} is thus equivalent to   
\begin{equation}\label{eq:normweigth}
    \begin{cases}
      w_k = \sum_{j=1}^n \Xi_{kj}^{-1};& k=1, \cdots, n\\ 
      \sum_{k=1}^n w_k = 1.
    \end{cases}
\end{equation}

Another important point is that the optimal GMV portfolio obtained from \ref{eq:normweigth} and \ref{eq:invert} does not depend on the scale of the eigenvalues $\lambda_k^*$ (or, equivalently, the average volatility). Thus, the only constraints we must impose on the eigenvalues is their non-negativity
\begin{equation}
    \zeta_k := \frac{1}{\lambda_k^*}\geq 0;\,\,\, \mbox{for}\,\,\, k=1, \cdots, n.
\end{equation}

Finally, the full QP problem is expressed as
\begin{equation}\label{eq:constrains}
    \begin{array}{ll@{}ll}
    \min_{{\bf w,\Xi^-1,\zeta}} &  \sum_{i=1}^n   w_i \Sigma^\textrm{out}_{ij} w_j & \\
    \text{subject to} & w_k = \sum_{j=1}^n \Xi_{kj}^{-1}; & k=1, \cdots, n\\ 
     & \sum_{k=1}^n w_k = 1\\
    & \Xi_{ij}^{-1} = \sum_{k=1}^n \zeta_k v_{ik}^\textrm{in}v_{jk}^\textrm{in}; & i,j = 1, \cdots, n\\
    & \zeta_k \geq 0; & k=1, \cdots, n.
    \end{array}
\end{equation}

Eq. \ref{eq:constrains} defines a convex Quadratic Programming problem, which can be solved by numerical methods.  In this QP problem formulation the variables ${\bf w}$ and ${\bf \Xi}^{-1}$ are slack variables that will be identified by the optimization algorithm. In total, the QP has $n(n+5)/2$ variables: $n$ for ${\bf w}$, $n(n+1)/2$ for ${\bf \Xi}^{-1}$, as the inverse covariance is symmetric, and $n$ for ${\bf \zeta}$ which are of interest here. The number of constraints are $(n^2+5n+2)/2$.
The resulting optimal $\zeta_k$ can be then normalized to have a set of eigenvalues whose sum equals expected volatility.                                                         

The procedure described above does not guarantee an ordered sequence of optimal eigenvalues. Let us therefore  add $n-1$  ordering constraints to Eq.\ \ref{eq:constrains}, which yields 
\begin{equation}\label{eq:constsort}
    \begin{array}{ll@{}ll}
    \min_{{\bf w,\Xi^{-1},\zeta}} &  \sum_{i=1}^n   w_i \Sigma^\textrm{out}_{ij} w_j & \\
    \text{subject to} & w_k = \sum_{j=1}^n \Xi_{kj}^{-1}; & k=1, \cdots, n\\ 
     & \sum_{k=1}^n w_k = 1\\
    & \Xi_{ij}^{-1} = \sum_{k=1}^n \zeta_k v_{ik}^\textrm{in}v_{jk}^\textrm{in};~ & i,j = 1, \cdots, n\\
    & \zeta_k \geq 0; & k=1, \cdots, n\\
    & \zeta_k \geq \zeta_{k-1}; & k=2, \cdots, n.
    \end{array}
\end{equation}

These constraints necessarily imply a less optimal solution with respect to the unsorted case. 
A Python implementation of both QP problems is available from \cite{package:optrie}.

\section{Results:  real Global Minimum Variance portfolios}
We apply both estimators to a data set of adjusted daily returns of the most capitalized US equities  spanning the  1995-2017 period.

The experiments are carried out in the following way. We randomly select two contiguous time intervals $[t-\delta_\textrm{in},t[$ and $[t,t+\delta_\textrm{out}[$ from the whole period,  remove all the stocks which have more than $20\%$ of missing values or zero returns, and discard any two stocks with an in-sample correlation larger than $0.95$. From the remaining assets, we randomly select $n=50$ stocks and we compute $\Sigma^\textrm{in}$ and $\Sigma^\textrm{out}$ from the two intervals respectively. 

From the eigenvector basis of $\Sigma^\textrm{in}$ and $\Sigma^\textrm{out}$ we compute the optimal eigenvalues (Eqs\ ~\eqref{eq:constrains}), the sorted optimal eigenvalue and the Oracle ones (Eqs\ \eqref{eq:constsort}) and the Oracle eigenvalues, which optimize the Frobenius norm. We stress that NLS aims to approximate the Oracle eigenvalues, thus produces slightly worse results \cite{bongiorno2022AO}. For that reason, we only include the Oracle eigenvalues in our study. 
Finally, we compute the out-of-sample (realized volatility) of GMV portfolios for each RIE.

\begin{figure}
    \centering
    \includegraphics[width=0.3\columnwidth]{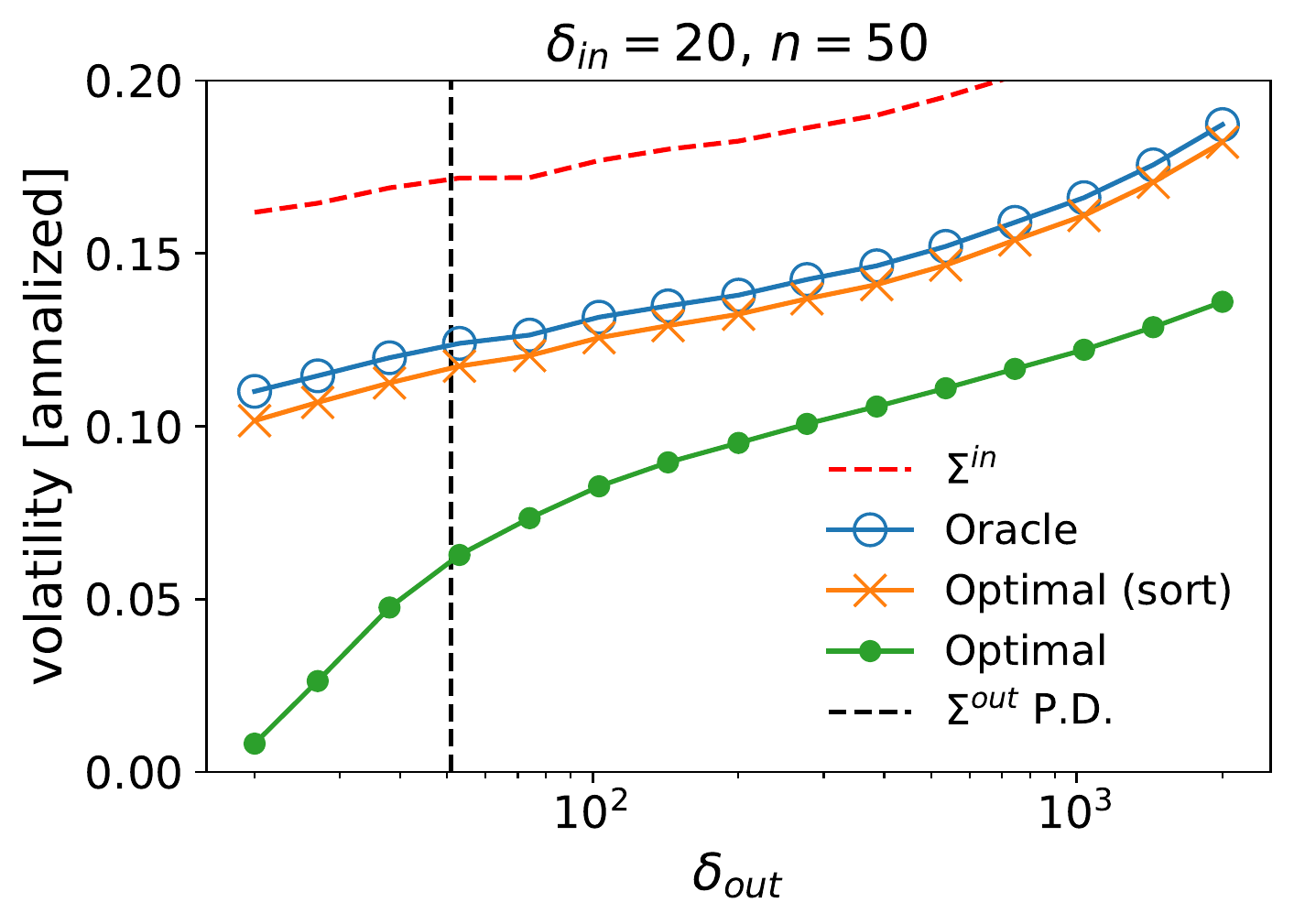}
    \includegraphics[width=0.3\columnwidth]{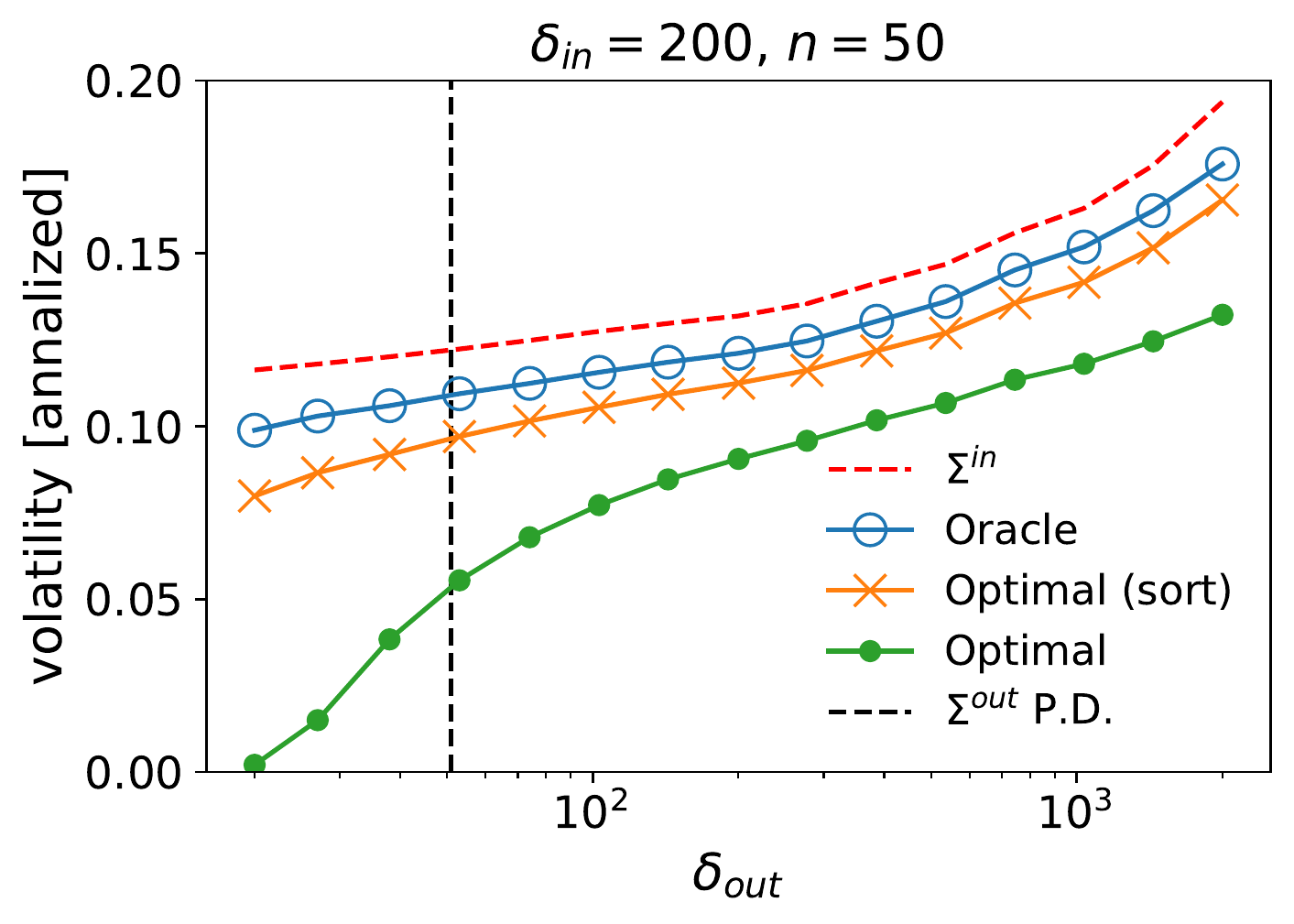}
    \includegraphics[width=0.3\columnwidth]{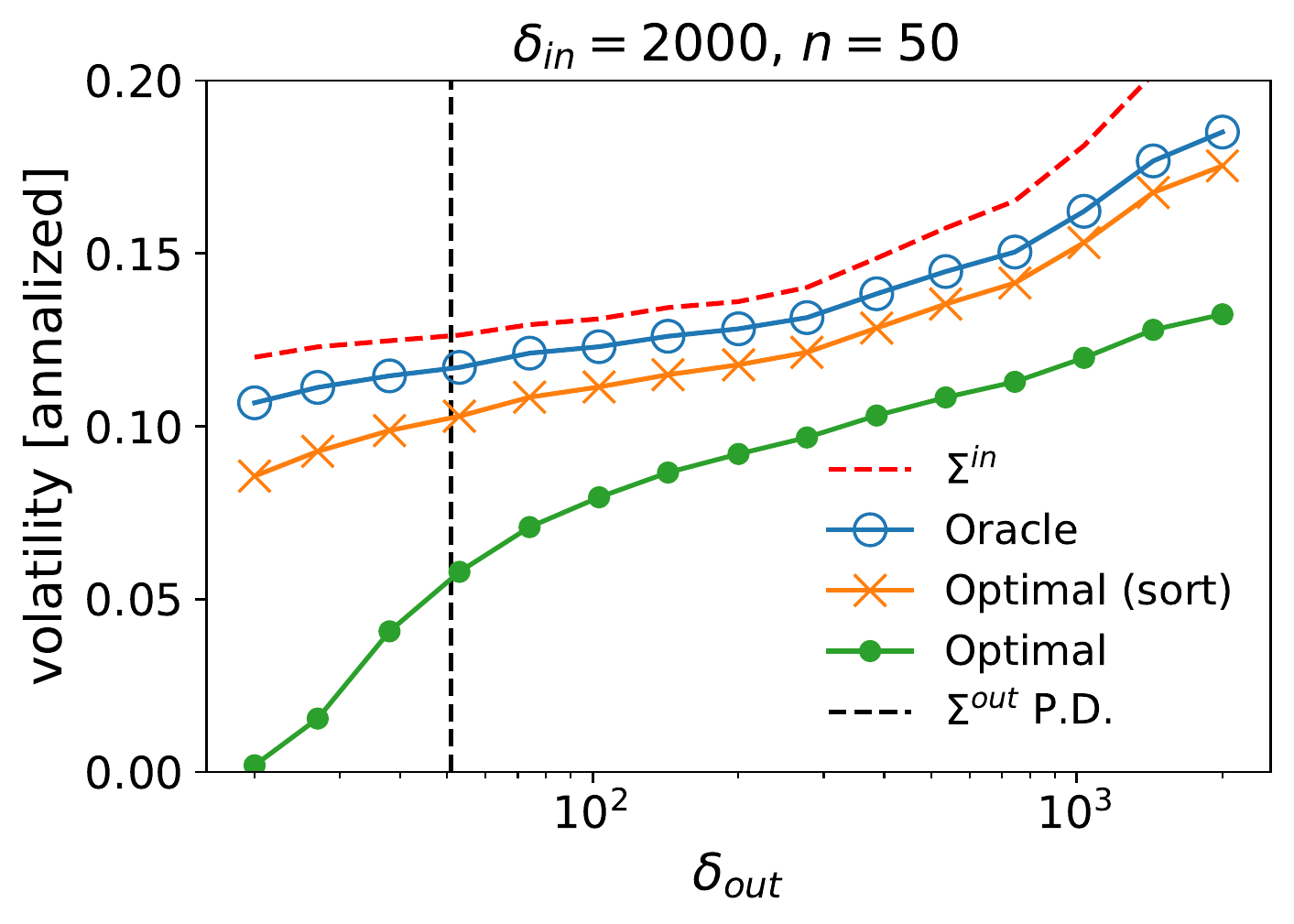}
    
    \includegraphics[width=0.3\columnwidth]{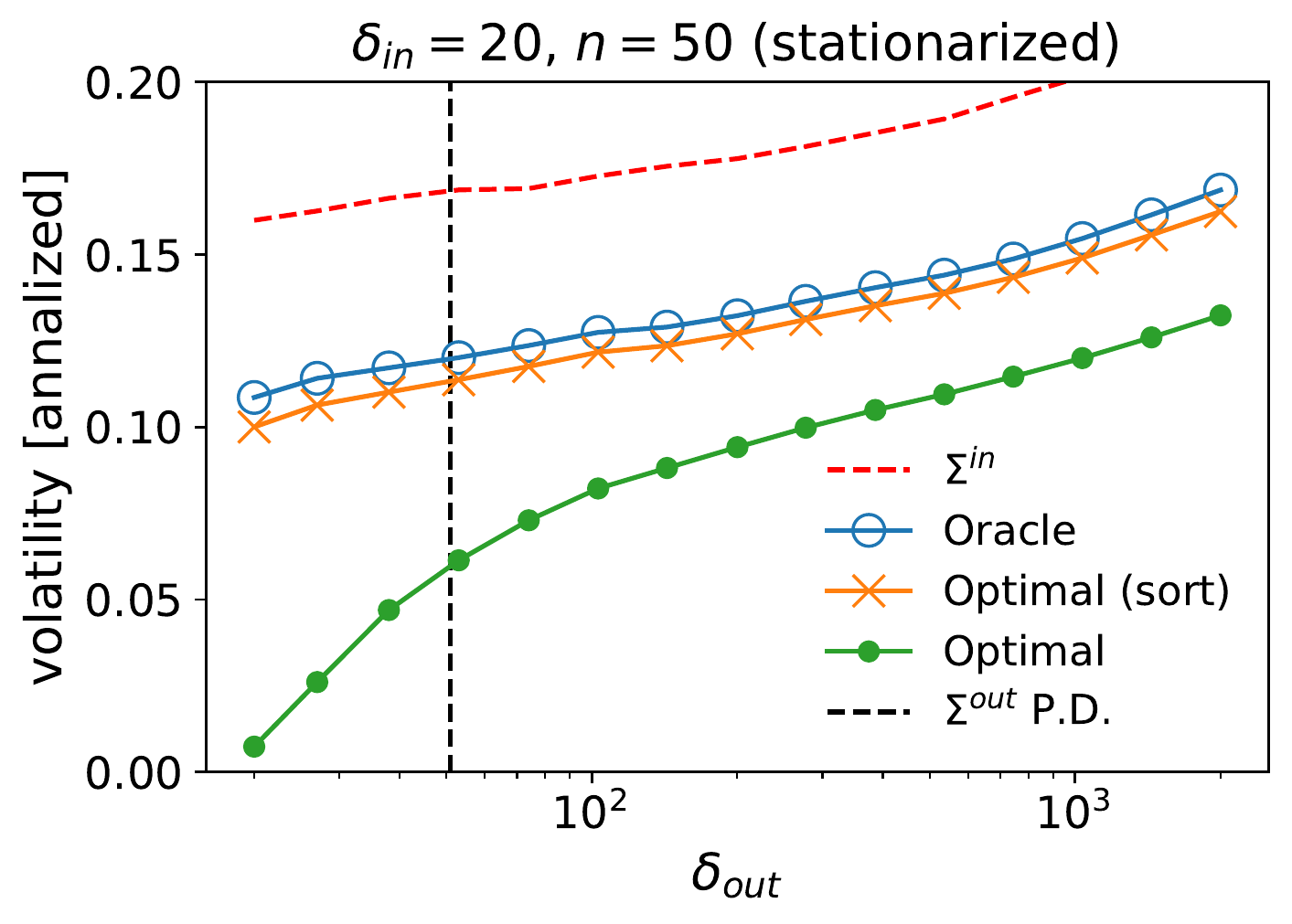}
    \includegraphics[width=0.3\columnwidth]{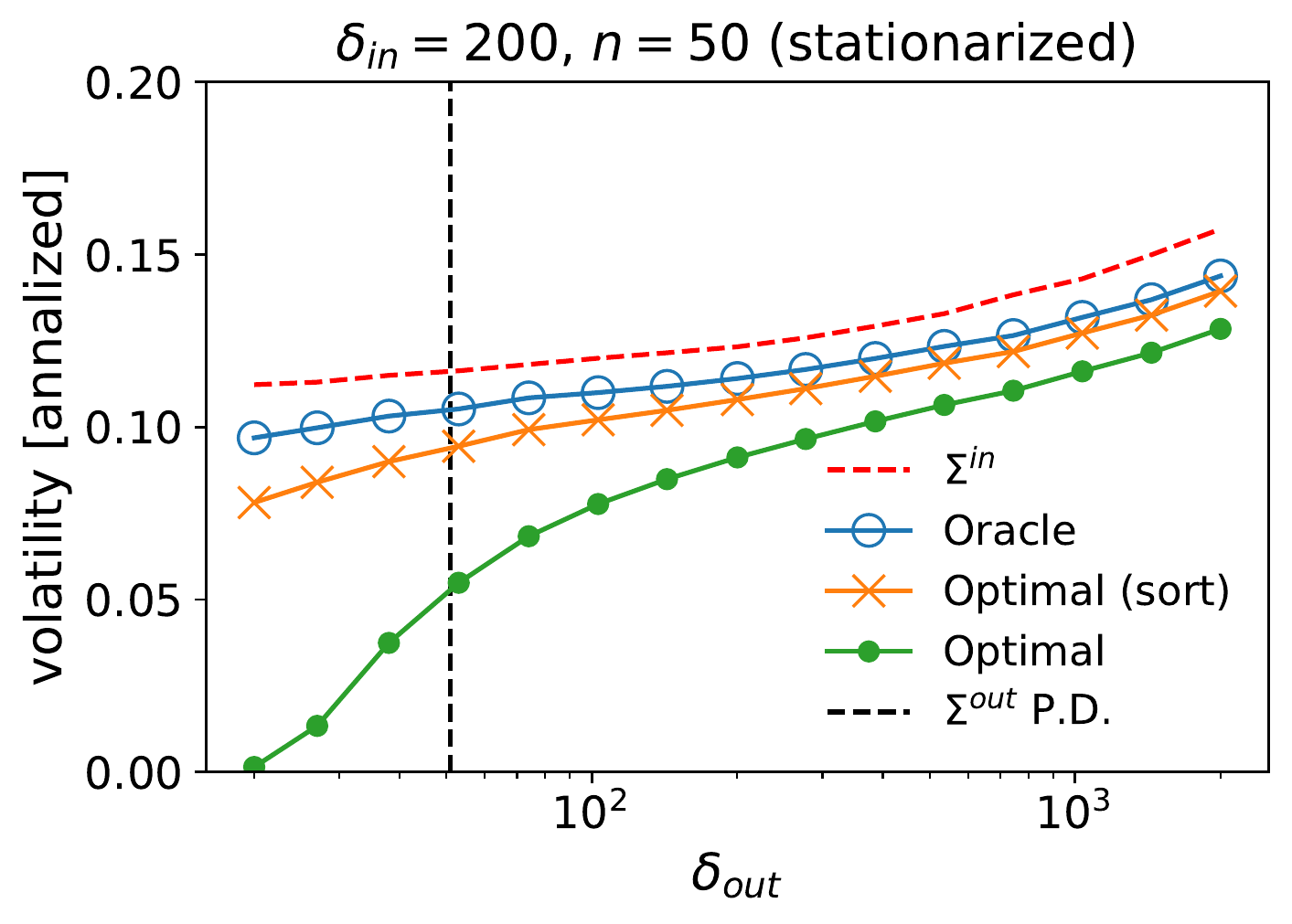}
    \includegraphics[width=0.3\columnwidth]{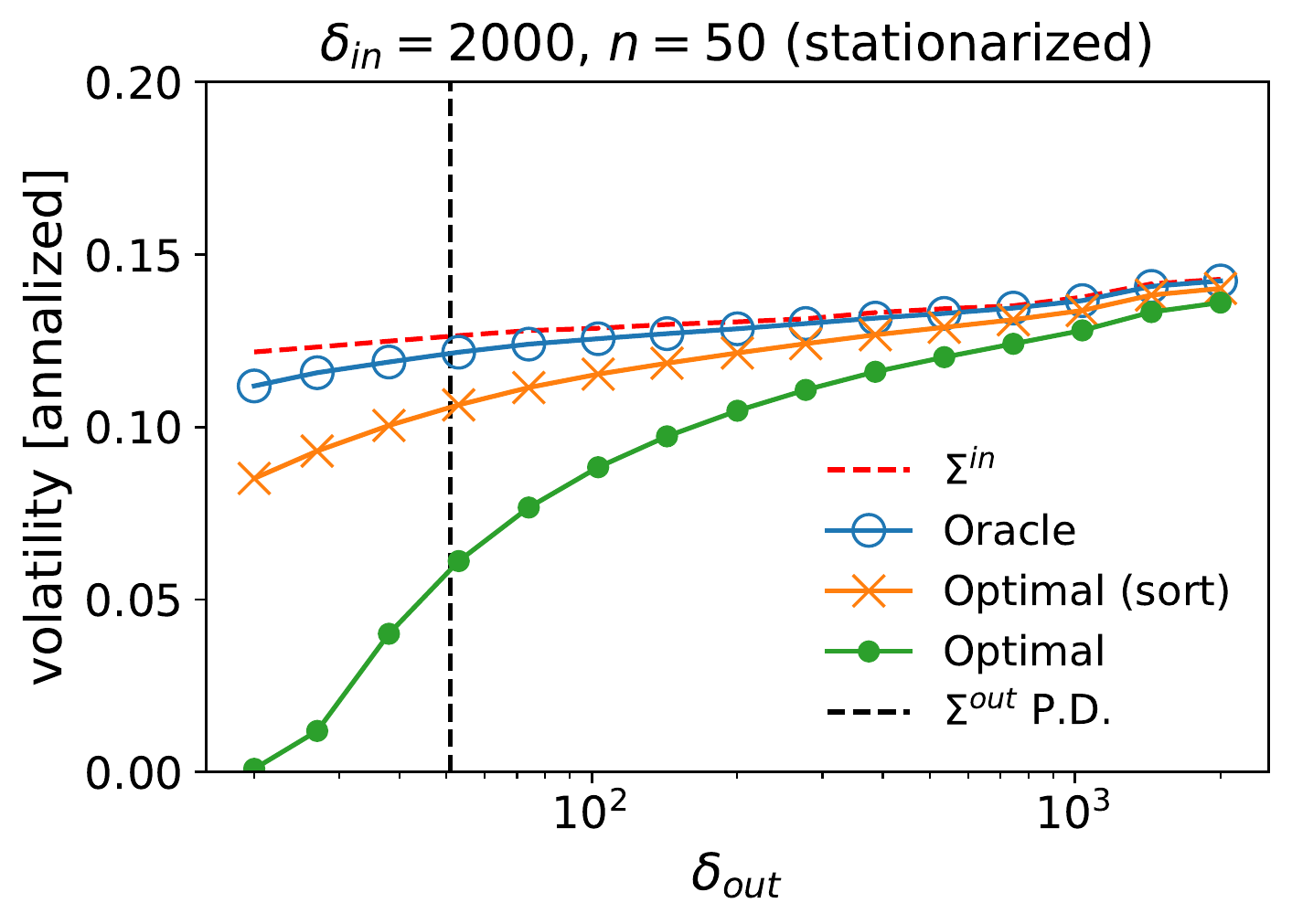}
    
    \caption{Realized volatility as a function of the out-of-sample period length. The upper panels display results from the original data, while the lower ones use stationarized data. The vertical black dotted lines refer to the transition point $n=\delta_\textrm{out}$ to a positive-defined out-of-sample covariance matrix. Matrix inversion of $\Sigma^\textrm{in}$ for $\delta_\textrm{out}<n+1$ is obtained with a pseudo-inverse~\cite{pantaleo2011improved}. Averages over 10,000 portfolios of $n=50$ random assets in random periods from daily data (large capitalization  US equities).}
    \label{fig:dtout}
\end{figure}

In the upper panels of Fig.~\ref{fig:dtout}, we show the average annualized volatility over  10,000  portfolios with randomly chosen assets at random times. The performance ranking is always the same one: applying Oracle correction is always better than using only the past ($\Sigma^\textrm{in}$, and the weights from the QP problem are always better than the Oracle RIE.

One notices a large gap between the Optimal and the Optimal sorted weights; when $\delta_\textrm{out}<n+1$, this gap comes from the fact that when $\delta_\textrm{out}<n+1$, the out-of-sample covariance matrix is not positively-defined. This means that there are $d = n+1-\delta_\textrm{out}$ null eigenvalues which imply an eigenspace of dimension $d$ from which every portfolio will have null variance. This is a purely mechanical effect that cannot be exploited from in-sample data only. This gap disappears when the monotonicity of the filtered eigenvalues is imposed. Interestingly, when $\delta_\textrm{out}>n+1$ the difference of performance between all the methods remains approximately constant until $\delta_\textrm{out}$ reaches very large values (1000 days, i.e., about 4 years of data). The phenomenon occurs when the in-sample window size $\delta_\textrm{in}$ increases from $200$ to $2000$. This effect comes from the non-stationarity of the dependence structure in the financial data we use. As a consequence, very large calibration or test periods produce out-of-date eigenvector bases.

To confirm this hypothesis, we stationarize the in-sample and out-of-sample time-series. The idea is simply to shuffle the in-sample and out-of-sample days of each subperiod  $[t-\delta_\textrm{in},t+\delta_\textrm{out}[$, in such a way that both the in-sample and out-of-sample holds a similar proportion of past and future days, which yields the same expected covariance matrix in both periods. In the lower panels of Fig.~\ref{fig:dtout} we show that increasing both $\delta_\textrm{in}$ and $\delta_\textrm{out}$ on two stationarized time-series reduces substantially the bias. This comes from the fact that in these conditions, the in-sample and out-of-sample eigenvector bases tend to be very similar. If either of the two time-window lengths are reduced, the similarity  between the two eigenvector bases decreases.

\section{Conclusion}

Covariance filtering for portfolio optimization, while having progressed much  in the last decades, still needs fundamental improvements. The problem lies in the nonstationary nature of dependence in financial markets, which conflicts with one of the main assumptions of the optimal stationary RIE, and which  implies that  the Frobenius distance is not 
the right cost function for the optimal RIE in a non-stationary world. 
The correct optimal RIE, derived in this work, currently does not have any asymptotical estimator that use only in-sample data, and thus, finding the optimal covariance cleaning scheme for equity markets is still an open question.

%From the application to US equities, one sees that the performance gap  between the Oracle estimator and the optimal correction decreases when the out-of-sample time-horizon is very large at fixed $n$. However, in practical applications from portfolio management the typical time-horizon ranges from a month to an year. 

Any improvement will mechanically improve on the state-of-the art DCC+NLS scheme \cite{engle2019largecov}. The simplest route is to keep improving RIEs for which many exact asymptotic results are known \cite{bun2017cleaning}. For example, we recently showed that a long-term average approach of the Oracle eigenvalues outperforms the Oracle eigenvalues in systems with nonstationary dependencies such as US and Hong Kong equity markets \cite{bongiorno2022AO}. 

We finally note that a RIE does not filter the noise in the eigenvectors which contain useful additional but noisy structures. A way to filter the latter is provided for example by ans\"atze such as hierarchical clustering \cite{tumminello2007hierarchically} or probabilistic hierarchical clustering \cite{bongiorno2021covariance,bongiorno2020reactive}, which outperform the optimal stationary RIE for GMV portfolios when $\delta_\textrm{in}<2n$ and DCC+NLS when the asset universe keeps changing, as it is the case in real life. An alternative route is to train a deep neural network to learn to predict both the eigenvalues and the eigenvectors.

%Although we do not know if the optimal eigenvalue correction could be useful for future analytical and machine-learning based approaches.

\section*{Acknowledgments}
This work was performed using HPC resources from the ``M\'esocentre'' computing center of CentraleSup\'elec and \'Ecole Normale Sup\'erieure Paris-Saclay supported by CNRS and R\'egion \^{I}le-de-France.
\section*{Funding}

This publication stems from a partnership between CentraleSup\'elec and BNP Paribas.
\bibliographystyle{unsrt} 
\bibliography{cas-refs}





% \end{thebibliography}
\end{document}